\documentclass[runningheads]{llncs}
\usepackage{graphicx}

\usepackage{cite}
\usepackage{amsmath,amssymb,amsfonts}
\usepackage{graphicx}
\usepackage{textcomp}
\usepackage{xcolor}
\usepackage{float}
\usepackage{comment}

\usepackage[spaces,hyphens]{url}
\usepackage[colorlinks,allcolors=blue]{hyperref} 

\begin{document}
\title{Voice Privacy with Smart Digital Assistants in Educational Settings\thanks{A version of this paper will appear in ITS 2021.}}
%
%

\author{Mohammad Niknazar\inst{1}\orcidID{0000-0001-8247-9331} \and
Aditya Vempaty\inst{1}\orcidID{0000-0001-9917-2985} \and
Ravi Kokku\inst{1}\orcidID{0000-0001-6418-1689}}
\authorrunning{M. Niknazar et al.}
%
\institute{Merlyn Mind, Inc., New York, NY, USA
\email{\{mohammad,aditya,ravi\}@xioresearch.com}}

\maketitle              
\begin{abstract}
The emergence of voice-assistant devices ushers in delightful user experiences not just on the smart home front, but also in diverse educational environments from classrooms to personalized-learning/tutoring. However, the use of voice as an interaction modality also could result in exposure of user's identity, and hinders the broader adoption of voice interfaces; this is especially important in environments where children are present and their voice privacy needs to be protected. To this end, building on state-of-the-art techniques proposed in the literature, we design and evaluate a practical and efficient framework for {\em voice privacy at the source}. The approach combines speaker identification (SID) and speech conversion methods to randomly disguise the identity of users right on the device that records the speech, while ensuring that the transformed utterances of users can still be successfully transcribed by Automatic Speech Recognition (ASR) solutions. We evaluate the ASR performance of the conversion in terms of word error rate and show the promise of this framework in preserving the content of the input speech. 


\end{abstract}
\section{Introduction}

There has been an explosion in voice-assistant device market over the past few years, especially due to the broader movement towards voice-enabling IoT devices in homes and enterprises \cite{CES2019}. This is due to the convenience of voice modality for human-machine interaction, compared to touch or typing. We envision that the momentum will promise to transform user experience in a variety of applications in the near future, including education. Even though voice-assistant devices in education are not meant to replace the meaningful and necessary human interaction between teachers and students, they can help teachers do many things more efficiently and excitingly. An example of this application would be a digital assistant that sits in the classroom to support the teacher (and sometimes students) in their everyday tasks. The teacher can activate it via voice whenever he/she wants assistance. This includes instant access to vetted content such as text, image, video, etc., or browsing web pages through voice.

However, the use of voice also raises practical privacy concerns~\cite{tech-crunch-privacy,cse-privacy,Liao2019}. These concerns are more serious in environments such as schools where children are present and they may also interact with the device. For instance, the speech files can be \textit{processed} for profiling after appropriate speaker identification (SID). Although recent privacy laws (including COPPA and GDPR) have forced major technology companies to establish privacy policies~\cite{voicepolicies}, a typical user is still unaware of and lacks enough forethought on ways in which the speech data could be used then or in the future (including sharing with third-parties). At the instant the user utters a command, the convenience of the service provided by a voice assistant takes more importance over the potential future uses of the collected data that violate privacy. Further, we believe that privacy should not be an after-thought, but should be provided by design~\cite{wiki-privacy}, and as close to the source of recording as possible.

Two main approaches could be considered to address the privacy problem described above. The first approach is to implement all aspects of speech intelligence, including automatic speech recognition (ASR) on the device locally. However, achieving cloud-comparable ASR performance using on-device ASR for broad vocabulary is an active research problem\cite{He2019}, and is also practically challenging due to the need for sophisticated hardware to support low latency requirements of voice assistants. 

The second approach, which is the focus of this paper, is to \textit{convert} audio signals before transmitting them to the cloud ASRs for transcription. The philosophy behind such an idea is that these voice assistants need to know \textit{what} was said, in order to function, but do not need information on \textit{who} said it. Hence, converting the audio signal so as to maintain the speech content while \textit{de-identifying} the speaker, serves our purpose. The goal then is to design a conversion system that ensures that speaker identification models on the cloud fail to associate the right speaker to an utterance, while keeping the ASR performance degradation within acceptable levels. More formally, given a speech dataset $\mathcal{D}$, and SID and ASR systems whose performance is characterized by $\mathrm{S}$ and $\mathrm{E}$ respectively, we want to design a speech conversion system $T_\Delta(\cdot)$ that satisfies the following:
\begin{eqnarray}
&\min_{T_\Delta}  \mathrm{S}[T_\Delta(\mathcal{D})] \nonumber\\
&\textrm{s.t. }  |\mathrm{E}[T_\Delta(\mathcal{D})]-\mathrm{E}[\mathcal{D}]|\leq \Delta.\label{eq:optimization}
\end{eqnarray}

\subsection{Related Work}
\label{related-work}

Different modalities of privacy protection including voice de-identification were discussed in \cite{ribaric2016identification}.
As one of the earliest attempts over thirty years ago, a text-dependent voice-conversion method based on vector quantization and spectral mapping was developed in \cite{abe1988s}, which had the limitation of using a parallel corpora for training. Various methods such as \cite{ney2004first} and  \cite{mouchtaris2004non} were then developed to remove this limitation and use non-parallel data for training, which is more practical for implementation. In addition to these methods that perform some kind of conversion from a source speaker to a target speaker, numerous methods (e.g. \cite{chaudhari2007privacy}), were developed to distort the voice by changing the pitch or some other basic characteristics of an input voice so that the resulted voice would sound synthetic and difficult to identify. A recent effort was carried out by Qian et al.~\cite{Qian2018}, in which a frequency warping function to the frequency domain signal was applied to change the voiceprint, and thereby hiding the identity. While there were promising results in this work, their method depends on a careful selection of the parameter values thereby reducing its applicability to different datasets with different acoustic conditions. 

Depending on the application, a reversible conversion may or may not be desirable. In \cite{jin2009voice}, a conversion based on Gaussian mixture model (GMM) mapping was used to convert voices of several source speakers to that of one synthetic target speaker. Having the conversion keys, one can revert the transformations and go back to the original voices. As the same authors showed in \cite{jin2009speaker}, increasing the number of speakers can lead to better de-identification and naturalness. Nevertheless, converting voices to synthetic voices has the drawback that if the system is hacked, the attacker will realize a conversion must have been performed on the voices as they sound synthetic, so they will try to revert the conversions by either hacking the system to access the conversion keys or even reverse engineering to estimate those keys. 

An approach that converts voices to synthetic ones and can still be very difficult or impossible to revert to the original voices was adopted in \cite{justin2015speaker}. In this approach, the input speech is first recognized with a diphone-based recognition system and converted into phonetic transcription. Then, a new synthetic speech is produced by a speech synthesis subsystem using the phonetic transcription. Although this approach offers a high level of protection of speaker identity, it requires an accurate diphone-based recognition system, which has its own challenges due to the limitations of such systems in different noise conditions and dealing with different accents. 

Another approach that can perform online voice de-identification and avoids some of the drawbacks of the above-mentioned methods was proposed in \cite{pobar2014online}. In this approach, a set of pre-calculated voice transformations based on GMM mapping is used to convert the voice of a new speaker to a synthetic voice. In order to select the best transformation, a speaker identification method is first applied to the incoming voice, then the appropriate transformation is selected. This approach which is inspired by approaches used for face de-identification (e.g. \textit{k-Same} \cite{newton2005preserving}), does not need a parallel corpus even for training of the initial set of transformations based on GMM mapping. However, in this approach, the target speaker was again a synthetic voice, which was trained using speech from four different male speakers. Therefore, if the system is hacked, the privacy can be compromised for the above-mentioned reasons.

\subsection{Contributions}
To address several drawbacks of the previous related work, we propose a framework that uses non-parallel data for training, with human sounding voice targets, which makes it more difficult to speculate if the original voices were transformed in the first place. Also, the targets for consecutive utterances are randomly selected to increase the robustness of the proposed method against reversing attempts. Moreover, the method can be applied to new users not seen in the training phase and with minimal/no need for dataset-specific tuning.

 In summary, we develop an efficient framework for voice privacy that is practical to implement in real-time by using a combination of speaker identification and speech conversion techniques, and developing a mapping approach for maximal speaker obfuscation with minimal impact on ASR accuracy. We demonstrate the efficacy of our framework with real-world voice utterances; our results show that we achieve obfuscation in a variety of source-target combinations while maintaining high ASR accuracy compared to the original speech.


The rest of the paper is organized as follows: Sec.~\ref{sec:methodology} describes the methodology used for de-identification to ensure privacy. Sec.~\ref{sec:exp} describes our experiments including data collection and results. Sec.~\ref{sec:conc} concludes the paper with some future directions.

\section{Methodology}
\label{sec:methodology}

\begin{figure}[t]
\begin{center}
\includegraphics[width=\textwidth]{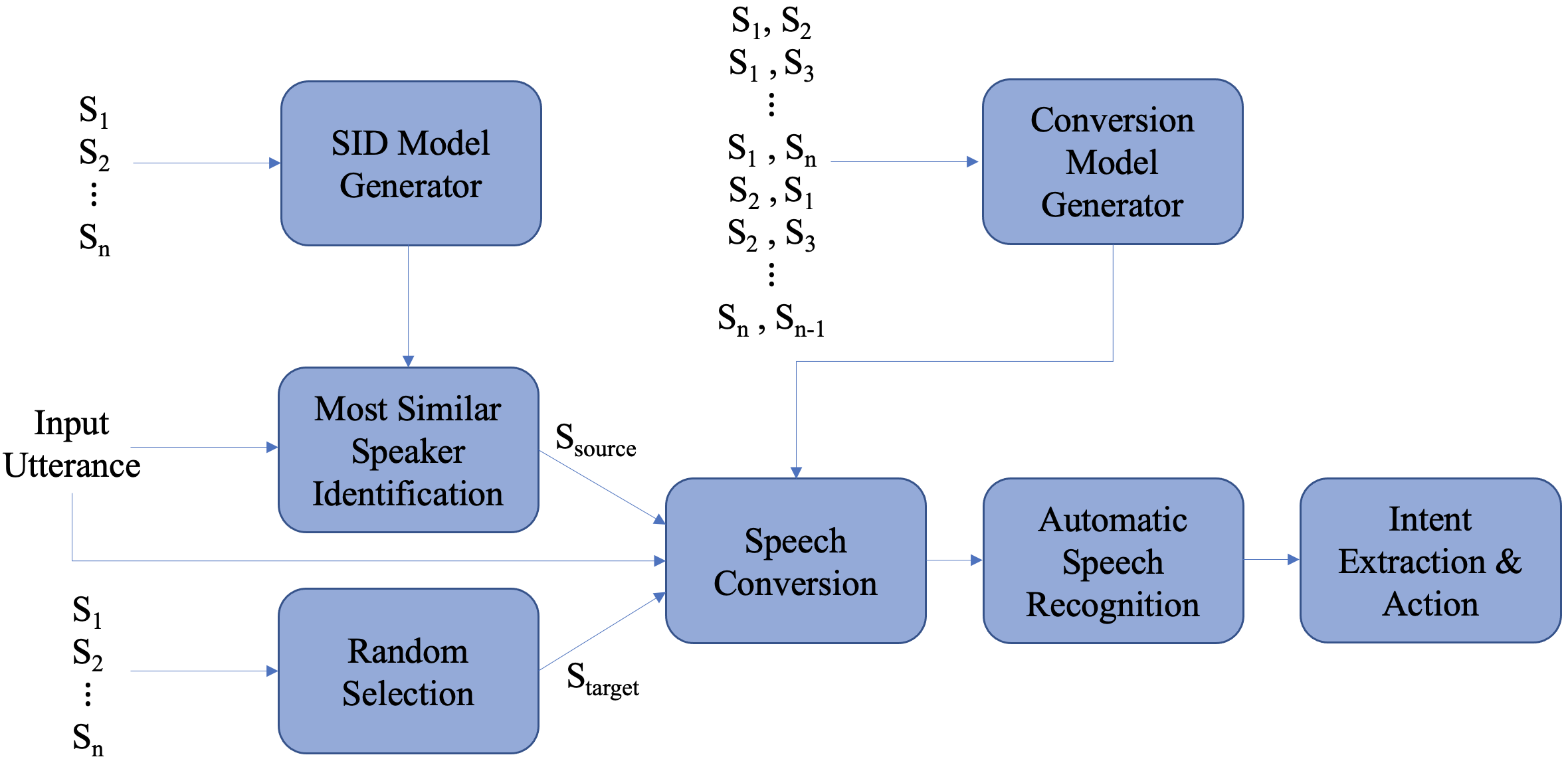}
\end{center}
  \caption{Block diagram of the proposed framework.}
  \label{Fig1}
\end{figure}

As represented by \eqref{eq:optimization}, our goal is to design a conversion algorithm $T_\Delta(\cdot)$ that significantly degrades the SID performance while limiting the degradation in ASR performance. In a preliminary work \cite{niknazar2019globalsip}, some early observations were reported for voice privacy protection based on cycle-consistent adversarial networks voice conversion (CycleGAN-VC) \cite{kaneko2017parallel}, which can be used as a potential candidate for $T_\Delta(\cdot)$ to convert speech utterances. 

CycleGAN-VC is a voice based extension of CycleGAN \cite{zhu2017unpaired} which was originally proposed for image-to-image translation from a source domain to a target domain using a training set of image pairs that were not necessarily aligned. This deep learning method uses conditional adversarial networks as a general-purpose solution to image-to-image translation problems. In addition to learning the mapping from input image to output image, these networks also learn a loss function to train this mapping. As a result, an important advantage of CycleGAN-VC is that it can learn a mapping from source to target speech without relying on time-aligned data. Therefore, it can be applied on non-parallel data. CycleGAN-VC2 \cite{kaneko2019CycleGAN-VC2} is an improved version of CycleGAN-VC that outperforms CycleGAN-VC in terms of naturalness and similarity for different speaker pairs, including intra-gender and inter-gender pairs \cite{kaneko2019CycleGAN-VC2}. We used CycleGAN-VC2 based on the implementation in \cite{CycleGAN-VC2-github} to create multiple source-target conversion models.



The steps of the proposed framework are as follows (refer to Fig.~\ref{Fig1}). First a training dataset with $n$ number of speakers is selected. This dataset is used to create $n$ SID models, one for each speaker, as well as $n(n-1)$ conversion models, two for each pair of the speakers (one per direction). For SID, we adopted a typical SID method that uses mel-frequency cepstral coefficient (MFCC) and linear predictive coding (LPC) features along with GMM and universal background model (UBM) based on the implementation in \cite{speaker-recognition-github} to model each speaker. As mentioned above, we used CycleGAN-VC2 to create all possible conversion models possible with the available dataset.

Then, for a given input utterance to the system, a combination of source and target is selected to perform the conversion. Different approaches may be pursued to select the conversion model for a given input utterance. One approach is to choose both source and target completely at random to maximize unpredictability of this selection. However, this approach assumes that all combinations of the source-target selections provide high-quality conversion, which may not necessarily be the case, as the pre-generated models may have different compatibility levels for different input speakers. On the other hand, if both source and target are selected in a deterministic manner, there will be a risk of hacking the system and reverting all the conversions. 

Therefore, in this paper, we use an approach which is a trade-off between the above-mentioned approaches to satisfy both high-quality model selection while ensuring there is some level of randomness in the system. The given input utterance is first mapped to the closest speaker among all or a subset of the speakers in the training dataset. For determining the closest speaker, the SID models trained for the speakers in the dataset are used. This step aims at choosing the most similar source to the input utterance to ensure a high-quality conversion. Then, the target is chosen at random from all the remaining speakers or a subset of the remaining speakers in the training dataset. Since part of the model selection process is random, it ensures there is no reversibility in identifying the speaker of any given utterance. Finally, the input utterance is converted using the pre-generated conversion models and the output is sent to the ASR and intent extraction modules to take an action. Figure \ref{Fig1} shows the block diagram of the proposed framework.

\section{Experiments}

\label{sec:exp}

\subsection{Setup}
In order to assess the performance of the proposed framework, 880 utterances were recorded from one woman and four men (176 each), where three men were native English speakers with American accent. The recorded utterances were typical examples of commands given to voice assistant systems in an educational setting such as \textit{“go back to the beginning of a video”}, \textit{“mute”}, \textit{“speak louder”}, \textit{“yes”}, etc. The data were recorded at 16 kHz using a miniDSP UMIK-1 microphone inside a quiet room. The data were further preprocessed to obtain mono \textit{‘.wav’} files at 16 bits. 

\subsection{Results}

 For the training dataset, VCC2016 dataset \cite{toda2016voice} was used. In this dataset, there are utterances from five male and five female subjects ($n$=10) that were recorded in a high-quality setting. These data went through two independent steps of model generation:
\begin{itemize}
	\item For each speaker, an SID model was created. This results in 10 models.
	\item For each pair of speakers as source and target, a conversion model was created. This results in 90 models.
\end{itemize}

Different combinations of source-target subsets were evaluated to identify the subset with the lowest Word Error Rate (WER). Table \ref{tbl1} reports the WER statistics for different subsets on the outputs obtained from a major technology vendor’s cloud ASR platform~\cite{google-cloud-asr} when an appropriate list of commands, which goes beyond the commands used by the users in the dataset was provided as the input context to the ASR. The p75 results for the WER show success of the conversion in most cases. As seen, the lowest WER was obtained when both male and female speakers were used as candidates to map the source, and only male speakers were used as the random target. This phenomenon may be caused by the inherent characteristics of the CycleGAN-VC2 method such that male targets that generally have lower voice fundamental frequencies lead to higher-quality outputs. Another possibility is that since the ASR engines were probably trained with datasets that have more male voices, these engines are generally more successful when the input utterance is from a male speaker than a female speaker. Therefore, for the remainder of the results, we focus on the conversion system where the \emph{source} is chosen to be the closest speaker from all 10 speakers in the training dataset, and the \emph{target} is selected randomly from the 5 male speakers.

\begin{table}[t]
\centering
\caption{WER comparison for different conversion combinations. The first row corresponds to the original utterances with no conversion. p75 Truncated Mean denotes the mean of the values up to the 75th percentile of the WER distribution. The median value for all cases was 0.}
\begin{tabular}{cccccc}

\hline

Source & Target & \begin{tabular}{@{}c@{}}p75 Truncated \\ Mean\end{tabular}   & Mean & STD \\
\hline
\_ & \_ & 0  & 0.019 & 0.107 &     \\
All & All & 0.028 & 0.219 & 0.363 &  \\
All & Male & 0.006 & 0.171 & 0.332 &  \\
All & Female & 0.044 & 0.256 & 0.389 &  \\
Male & Male & 0.010 & 0.186 & 0.342 &  \\
Female & Female & 0.016 & 0.193 & 0.346 &  \\
Male & Female & 0.158 & 0.369 & 0.436 &  \\
Female & Male & 0.032 & 0.235 & 0.378 & \\
\hline
\end{tabular}
\label{tbl1}
\end{table}

In order to evaluate the de-identification performance of the proposed method, different approaches may be adopted. In a real-world scenario, the cloud speech processing systems will not necessarily already have the labels for each speaker to train a model for the given speaker and use it later to identify that person. Hence, they need to address the speaker identification task with an unsupervised method whose performance is expected to improve as more data are received. Moreover, these engines typically receive the original voice from each speaker, not randomized converted voices which may sound significantly different from the original voice. Nevertheless, to simplify things in favor of a cloud SID system aiming at identifying speakers and potentially compromise their privacy, we assumed that the system already has some training data on original voices of each speaker with labels and generated SID models based on them (same SID approach as discussed in Sec.~\ref{sec:methodology}). Then, we compared the performance of SID with unconverted and converted test data. In order to do so, we randomly divided the data of 880 utterances to train and test sets with 70\% and 30\% portions, respectively. The SID on the original unconverted test data was 100\% accurate with all utterances being perfectly classified. However, when the converted version of the same utterances were tested with the SID, the accuracy was only 20.45\% with 210 out of 264 utterances mis-classified (essentially random selection of speaker). This shows the effectiveness of the proposed framework in disguising the identity of the speakers.

\begin{figure}[t]
  \includegraphics[width=\linewidth]{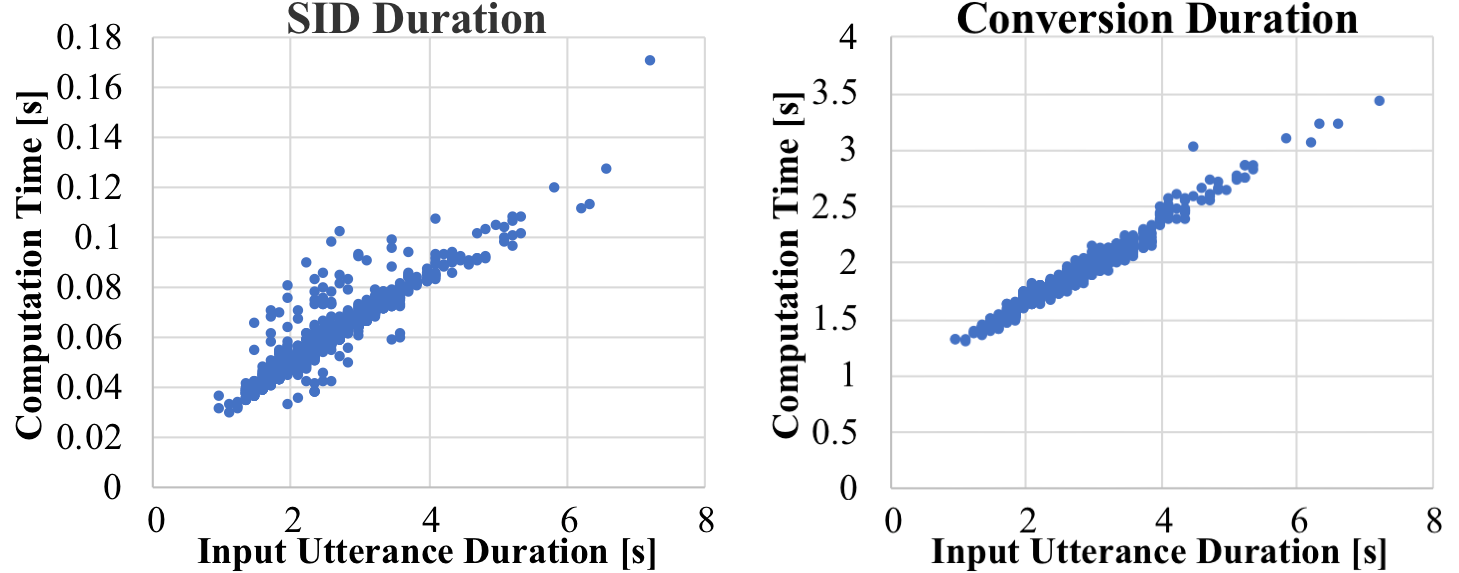}
  \caption{Computational time of the proposed framework for speaker identification (SID) and voice conversion.}
  \label{Fig3}
\end{figure}

Finally, the inference computation time of the proposed framework was calculated for the input utterances to determine the feasibility of the method in real-time applications. Fig.~\ref{Fig3} shows the computation time for speaker identification and voice conversion on a machine with an Nvidia GPU (1080 GTX), with less than 10\% GPU utilization.
As seen, speaker identification is performed very fast and remains under ~0.2s even for an utterance of ~7s. Voice conversion takes longer and a delay of ~0.5s is observed for short utterances of ~1s, which can be negligible in most applications. Nevertheless, as the duration of utterances increases, the delays become smaller and for any utterances of 2s and longer, the conversion can be run real time.

\section{Conclusion}
\label{sec:conc}
In this paper, we proposed a practical framework for voice privacy protection in educational settings, while preserving the content of an utterance, using a combination of speaker identification and speech conversion models. While the techniques are equally applicable in general home settings, voice assistants in homes seem to have found their place already even without strict privacy controls.  In educational settings, however, due to laws and regulations, solutions may need extra level of privacy before gaining adoption. 
We believe that if smart assistants provide such voice privacy by design (at the source), the adoption of voice interfaces would accelerate in both one-one personal tutoring and public spaces such as schools and lead to ubiquity of voice as an interaction modality sooner. Further, we believe that solutions like the one proposed in this paper let developers of educational solutions continue to use state of the art cloud ASR solutions provided by third-party organizations that maximize speech recognition accuracy.

%
%
%
%

\bibliographystyle{splncs04}
\bibliography{main}

\begin{thebibliography}{10}
\bibitem{tech-crunch-privacy}
TechCrunch: {}41\% of voice assistant users have concerns about trust and
  privacy
\end{thebibliography}

\end{document}